\documentclass[epj]{svjour}
\usepackage{graphics}

\begin{document}
\title{Deviation of the Nucleon Shape From Spherical Symmetry: Experimental Status}
%\footnote{Invited talk at Electron-Nucleus Scattering VII, June 23-28,2002, Elba, Italy} 
\author{A.M.Bernstein\thanks{Expanded version of an Invited talk at Electron-Nucleus Scattering VII, June 23-28,2002, Elba, Italy.}}
\institute{M.I.T., Cambridge, Mass., U.S.A.}

\date{\today}
\abstract{In this brief pedagogical overview the physical basis of the deviation of the nucleon shape from spherical symmetry will be presented along with the experimental methods used to determine it by the $\gamma^{*} p \rightarrow \Delta$ reaction.The fact that significant non-spherical electric(E2) and Coulomb quadrupole(C2) amplitudes have been observed will be demonstrated. These multipoles for the $N,\Delta$ system as a function of $Q^{2}$ from the photon point through 4 $GeV^{2}$ have been measured with modest precision. Their precise magnitude remains model dependent due to the contributions of the background amplitudes, although rapid progress is being made to reduce these uncertainties. A discussion of what is required to perform a model independent analysis is presented. All of the data to date are consistent with a prolate shape for the proton (larger at the poles) and an oblate shape(flatter at the poles) for the $\Delta$. It is suggested here that the fundamental reason for this lies in the spontaneous breaking of chiral symmetry in QCD and  the resulting, long range(low $Q^{2}$), effects of the pion cloud. This verification of this suggestion, as well as a more accurate measurement of the deviation from spherical symmetry, requires further experimental and theoretical effort.}

\maketitle
\section{Introduction}
\label{intro}
Experimental confirmation of the deviation of the proton structure from
spherical symmetry is fundamental and has been the subject
of intense experimental and theoretical interest\cite{Nstar}
since this possibility was originally raised by
Glashow\cite{Glashow}. The most direct method to determine this would be to measure the quadrupole moment of the proton. However since the proton spin is 
1/2 this is not possible. Therefore, this
determination has focused on the measurement of the electric
and Coulomb quadrupole amplitudes (E2, C2) in the
predominantly M1 (magnetic dipole -quark spin flip)
$\gamma^{*} N \rightarrow \Delta (1/2 \rightarrow 3/2) $ transition. Thus measurements of the E2 and C2 amplitudes represent deviations from spherical symmetry of the $N,\Delta$ system and not the nucleon alone. The experimental difficulty is that the  E2/M1 and C2/M1 ratios are small (typically $\simeq$ $-2$ to
$-8$ \% at low $Q^{2}$). In this case the non-resonant
(background) and resonant quadrupole amplitudes are the same order of
magnitude. For this reason experiments have to be designed to attain the
required sensitivity and precision to separate the signal and background contributions.
This has been accomplished for photo-pion reactions for the E2 amplitude using polarized photon beams\cite{Mainz,LEGS}. In pion electroproduction the deviation from spherical symmetry is easier to observe due to the interference between the longitudinal(Coulomb) C2  and the dominant M1 amplitudes by observation of the $\sigma_{TL}$ cross section\cite{Mertz}. Electroproduction experiments are also being performed for a range of four momenta $Q^{2}$\cite{Mertz,warren,Kunz,Mpolp,MainzTL',Bonn,Joo,Stoler,HallA} which provides a measure of the spatial distribution of the transition densities. On the other hand the presence of the additional longitudinal multipoles in electroproduction means that there are more observables to measure and therefore more data must be taken than in photoproduction experiments. The experiments to generate an extensive data-base that would allow a model independent analysis have just begun, both in photo- and in electro-production. At the present time one must rely on reaction models to extract the resonant  M1, E2, and C2 amplitudes of interest from the data. As has been pointed out  the model
error can be much larger than the experimental error\cite{Mertz}. 
Therefore it is important to test model calculations for a
range of center of mass(CM) energies $W$ in the region of 1232 MeV, the
resonant energy, which provides a range of the relative
background and resonant amplitudes, as well as picking out specific
observables which are sensitive to the quadrupole amplitudes (e.g. $\sigma_{\rm TL}$) and others which are primarily sensitive to the
background amplitudes(e.g. $\sigma_{\rm TL^{'}}$).

\section{Why Should the Nucleon be Deformed?}
\label{sec:1}
\begin{figure}
\begin{center}
%\makebox{5cm, 5cm}[pos]{(l)}
%[htbp]
\resizebox{0.25\textwidth}{!}
%\vspace{5cm}
{\includegraphics{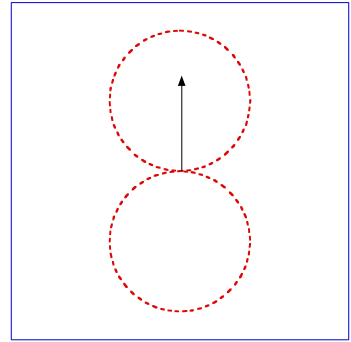}}
\end{center}
\caption{The pion cloud contribution to nucleon structure. The arrow represents the nucleon spin vector and the dotted line the pion cloud.}
\label{Npion}
\end{figure}

It is well known that in the quark model there are non-central (tensor) interactions between quarks which were modeled after the electromagnetic interaction\cite{Glashow,Isgur}.\\ Although this interaction does, in fact, introduce non-spherical responses (E2 and C2) into the electromagnetic matrix elements they are only a small fraction of the observed amplitudes. In my view this is not surprising since the long distance part of the nucleon and $\Delta$ structure should be related to the pion cloud which is poorly represented in quark models. We expect the long range (low $Q^{2}$) behavior to be pion field  dominated since it is the lightest  hadron. Of course this is well known experimentally and is a cornerstone of classical nuclear theory. What is new is our more recent understanding  that the pion itself, and its interaction with other hadrons, is a consequence of  spontaneous chiral symmetry breaking in QCD\cite{chiral}. In the chiral limit, i.e. where the light quark masses are set equal to zero, the QCD Lagrangian has chiral symmetry which does not appear in nature. We know this experimentally since if chiral symmetry were exact we would observe parity doubling of all hadronic states. This means that the chiral symmetry is broken (or more exactly hidden) and is manifested in the appearance of zero mass, pseudoscalar Goldstone Bosons. Since, in nature, the light quark masses are small but non-zero, the physical Goldstone Bosons have a small mass and are identified as the $\pi$ mesons. The coupling of a Goldstone Boson to a nucleon is g$\vec{\sigma}\cdot \vec{p}$ where g is the $\pi-N$ coupling constant (predicted by the 
Goldberger-Trieman relation) $\sigma$ is the nucleon spin, and p is the pion momentum. This interaction vanishes in the s wave and leads to the Goldstone theorem that the interaction vanishes as $p \rightarrow 0$. The $\vec{\sigma}\cdot\vec{p}$ interaction is strong in the p wave which leads to the $\Delta$ resonance and is the basis of the deviation from spherical symmetry in the nucleon, illustrated schematically in Fig.~\ref{Npion}, and $\Delta$ structure. In a sense this is the basis of classical nuclear theory. 

Although it is beyond the scope of this presentation, I cannot resist mentioning that the $\vec{\sigma}\cdot\vec{p}$ interaction required by the spontaneous breakdown of chiral symmetry in QCD economically describes the basics of $\pi N$ scattering and of nuclear physics. Indeed, this $\pi N$  interaction leads to a  non-spherical NN interaction (the tensor force). By generalizing Fig.~\ref{Npion} to the case of two nucleons, it is not hard to see physically (semi-classically) that the most attractive position for two nucleons is for one nucleon to be spatially above the second with their spins parallel. This is the configuration which is favored by the tensor force. 

Quantitatively the quark model calculations of the M1 matrix element are  too small\cite{Capstick}(although this is not often discussed). This is shown in Table 1 where the magnitude of the experimental\cite{PDG} and theoretical matrix elements for the $\gamma N \rightarrow \Delta$ reaction are presented. It can be seen that the quark model predictions are $\simeq 30\%$ too low for the dominant M1 and an order of magnitude too small for the E2 matrix element, which is the indicator of the non-spherical structure of the nucleon and $\Delta$ structure. The table also includes the redundant E2/M1 ratio just to illustrate that it has become commonplace in the recent literature to quote only this latter ratio and not the absolute values of the matrix elements. By doing so, some important lessons tend to be overlooked. As an example, we focus on the quark model extensions\cite{Buchmann}, which introduce multi-body interactions between the quarks, taking into account the composite nature of the constituent quarks. These currents take the pion field partially into account and, as can be seen in Table 1, this effect increases the E2 matrix element to the empirical size and in fact increases the E2/M1 ratio to an even larger value than the experiment. However this treatment does not  increase the magnitude of the M1 matrix element so it remains $\simeq 30\%$ less than experiment. As was discussed above, based on spontaneous chiral symmetry breaking, it is physically intuitive that this shortfall should be looked for in the long range effect of the pion cloud.

This issue of the quark core and pion cloud contributions has been addressed in a meson exchange model by Sato and Lee\cite{SL}. Their model results are also presented in Table 1. Here it is seen that their model tends to make up for the deficiencies of the quark model, not only for the E2/M1 ratio, but for the individual magnitudes of the E2 and M1 matrix elements. The uncertainties shown in Table 1 are due to uncertainties in their model parameters (see Table IV and the discussion in their 1996 paper\cite{SL}). Sato and Lee have also calculated the effects of the pion cloud for pion electroproduction as a function of $Q^{2}$ and the results are presented in Fig. 2. It can be seen that the enhancement of the M1 matrix element is significant and that the non-spherical E2 and C2(Coulomb quadrupole) matrix elements are dominated by meson cloud effects. It is also seen in Fig. 2 that, as expected, the long range pion cloud effects are more dominant at low $Q^{2}$. The dynamic Sato-Lee calculations are in excellent agreement with the data for photo-pion production in the $\Delta$ region (some of the parameters were fit to these data) and are also in good agreement with the JLab data presented by Burkert at this meeting. 
However, before becoming complacent, we should note that  the next section will show that this model is not in agreement with low $Q^{2}$ data taken at Bates\cite{Mertz,Kunz} near the predicted peak of the pion cloud contribution.This suggests to me  that even though the Sato-Lee Lagrangian has chiral symmetry they are not completely implementing the dynamics of chiral symmetry breaking, perhaps in their omission of the pion loops which are required in chiral perturbation theory calculations of the $e p \rightarrow e \pi^{0}p$ reaction in the near threshold\cite{BKM} and $\Delta$ regions\cite{Hemmert}. Therefore, even though the contribution of the pion cloud appears to be required, its description must be considered somewhat qualitative at this point. It's detailed contribution needs to be calculated in a manner which is more consistent with the dynamics of spontaneous breaking.

Calculations of the $\gamma^{*} N \rightarrow \Delta$ transition in chiral perturbation theory (ChPT), which incorporates spontaneous chiral symmetry breaking in a manner which is consistent with QCD, have just begun\cite{Hemmert}. In this calculation the $\Delta$ is included as a dynamical degree of freedom. A low energy expansion has been performed in the small parameter 
$\epsilon =( p_{\pi},m_{\pi},\delta= M_{\Delta}-M_{p})$ 
 to order $O(\epsilon^{3})$.In this calculation the importance of the pion loops is clear since they are required by the power counting rules of ChPT at $O(\epsilon^{3})$, or higher. As is typical of low energy effective theories there are two low energy parameters which were fit to the empirical M1 and E2 transitions so that there aren't any predicted values to be included in Table 1. The slopes of the transition form factors are predicted, but cannot yet be compared to experiment since there is a paucity of low $Q^{2}$ data, which is something that we plan to remedy in the next year. On the theoretical side it is important that the ChPT calculations be carried out to make contact with experimental observables and also extended to $O(\epsilon^{4})$ to check the convergence of these calculations.  
It is clear that at the present time the verification of the idea that the pion cloud is the major contributor to the nucleon and $\Delta$ deformation requires more theoretical and experimental work.This is important, since as was stated above, it is expected that the pion cloud plays a significant role in baryon structure\cite{Ulf}. 

The question of the shape of the nucleon and $\Delta$ were explored in the context of three different models by Buchmann and Henley\cite{BH}. They conclude that the proton is prolate (longer at the poles) and the $\Delta$ is oblate(flatter at the poles). This is consistent with the  data. 
\begin{table}
\begin{center}
\caption{Experimental and Model Amplitudes for the $\gamma N \rightarrow \Delta$ Reaction.The units for M1 and E2 are $10^{-3} GeV^{-1/2}$ and E2/M1 in \% The numbers in parenthesis are experimental or model errors}
\label{tab:1}     
\begin{tabular}{{|c|c|c|c|}}\hline
model& M1 & E2& E2/M1\\
Experiment\cite{PDG}&288(8)&-7.2(0.5)&-2.5(0.5)\\
QM:Capstick\cite{Capstick} &196 & -0.1&-0.04\\
QM:Buchmann\cite{Buchmann}&203&-7.0&-3.5\\
%Linear Sigma&226&-3.9&-1.7\\
%Cloudy Bag&288&0.1&0.04\\
SL(bare)\cite{SL}&175& 0.0(2.3)&0.0(1.3)\\
SL(dressed)\cite{SL}&258&-4.6(2.3)&-1.8(0.9)\\
\hline
\end{tabular}
\end{center}
\end{table}
\begin{figure}
\begin{center}
\resizebox{0.2\textwidth}{!}
%\vspace{5cm}
{\includegraphics{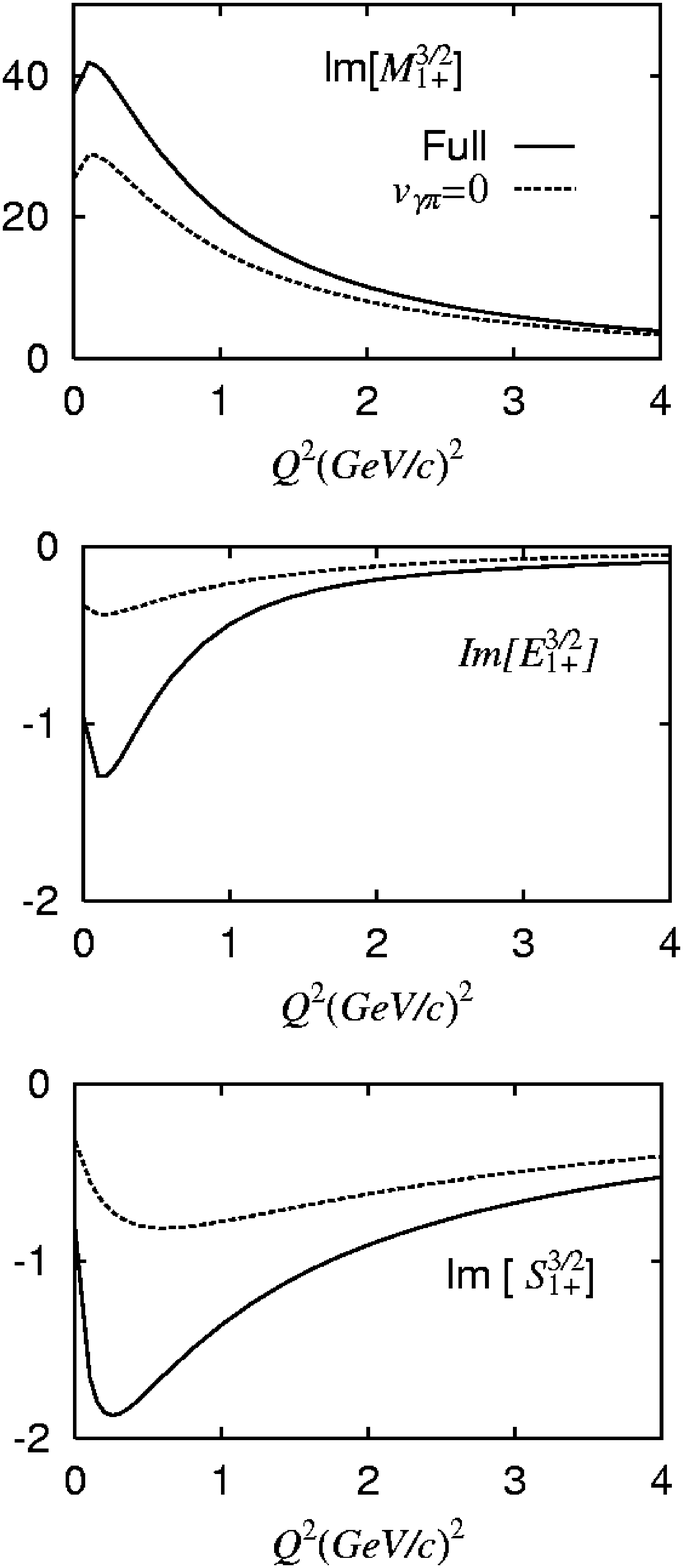}}
\end{center}
\caption{The contribution of the quarks and pion cloud to the M1, C2, and E2 transition amplitudes in the $\gamma^{*} p \rightarrow \Delta$ reaction calculated by Sato and Lee\cite{SL}}
\label{pioncloud}
\end{figure}

Finally we note that there are two lattice QCD calculations\cite{lattice}at $Q^{2}=0, 0.52GeV^{2}$ which demonstrate non-zero quadrupole transition amplitudes. Taking into account the relatively large theoretical errors, as well as the experimental errors, they are tolerably close to the data. Clearly we await further calculations which reduce the error, so that accurate QCD calculations can be compared to experimental results. We do note however that the question of exactly how to compare the experimental transition amplitudes, which contain background continuum contributions, and a theoretical calculation which assumes that the $\Delta$ is a bound particle, need improvement.

\section{ Experiments on Proton Deformation}
\label{sec:2}

Modern photon experiments have been carried out at Mainz \cite{Mainz} and Brookhaven\cite{LEGS} of the $\vec{\gamma} p \rightarrow \pi^{0} p$ and $\vec{\gamma} p \rightarrow \pi^{+} n$  reactions with polarized photons. The combination of accurate measurements and the use of polarized photons provides sufficient sensitivity so that two laboratories have reported measurements of both the  small E2 amplitude and the dominant M1 amplitude. The observation of both charge channels allows an isospin separation of the resonant $\it{I}=$3/2 channel. The results for the polarized photon asymmetry are presented in Fig. 3. There is good agreement for this quantity between the two labs and model calculations and the results are $E2/M1 = -2.5 \pm 0.5\%$ \cite{PDG} showing that there is significant deformation in the $N,\Delta$ system (see the discussion in\cite{LEGS}). It should be mentioned, however, that although there is very good agreement between the Mainz and Brookhaven measurements of the polarized photon asymmetries shown in Fig. 3, there is a significant deviation in the unpolarized cross sections\cite{Mainz,LEGS} which unfortunately is still unresolved.  
\begin{figure}
\resizebox{0.5\textwidth}{!}
%\vspace{5cm}
{\includegraphics{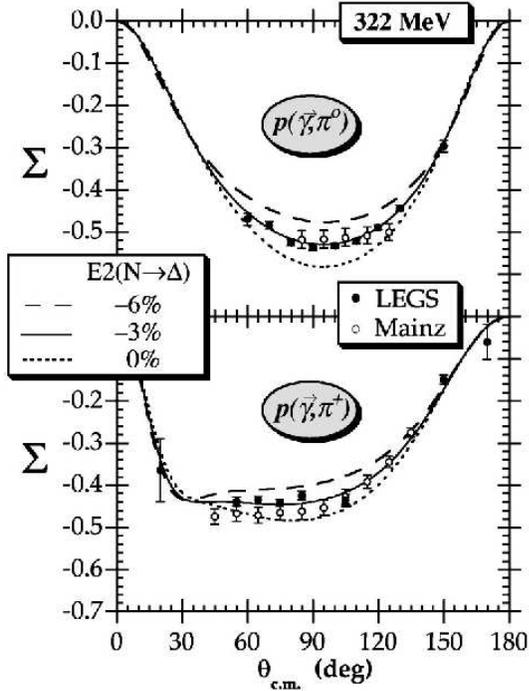}}
\caption{Polarized photon asymmetries for the $\vec{\gamma} p \rightarrow \pi^{0} p$ and $\vec{\gamma} p \rightarrow \pi^{+} n$  reactions plotted versus $\theta$. 
The curves are MAID\cite{MAID} for different E2/M1 ratios as shown.}
\label{polphoton}
\end{figure}

The situation for pion electroproduction in the $\Delta$ resonance region is evolving rapidly with activity at all the intermediate energy facilities: Bates\cite{Mertz,warren,Kunz}, Mainz\cite{Mpolp,MainzTL'}, Bonn\cite{Bonn}, and JLab\cite{Joo,Stoler,HallA} (for a review see\cite{Nstar}). In this brief report only a few highlights will be mentioned. The work at Bates will be emphasized here since the focus of this talk is on the pion cloud effects which are largest in the low $Q^{2}$ regime covered by those experiments and also since the JLab work was covered by Burkert at this meeting\cite{VB}. There has been a corresponding increase in the theoretical activity in this field which is being emphasized in a complementary talk at this meeting by Tiator who also presents an overview of the data\cite{Tiator}. 
The goal of the pion electroproduction experiments is to obtain accurate data which have sufficient sensitivity to determine the M1, E2, and C2 resonance amplitudes, but also sufficient coverage to determine the background amplitudes which are of the same order of magnitude as E2 and C2. These are of interest in their own right as an integral part of the $\gamma \pi N$ system and therefore as part of the chiral structure of matter. For example, the resonance and background amplitudes are related to the form factors and to the electric and magnetic polarizabilities of the nucleon. The background amplitudes contribute to the cross sections linearly with the resonance amplitudes, and the interference terms therefore make significant contributions to the observables. The background contributions also occur in the resonance amplitudes and are part of the physics. In a sense this is a crucial difference between dynamic and static models(e.g.the quark model) for which the $\Delta$ is treated as a bound state, which ignores the background contributions. The cleanest experimental determination would consist of an empirical multipole analysis of the data. At the present time we do not have a sufficient, accurate data-base with which to perform such an analysis and must rely on empirical models to extract the resonant amplitudes. The present short term experimental goal is to provide a sufficiently sensitive and accurate data-base to rigorously test the models. It is hoped that the combination of Born terms and the tails of higher resonances will suffice to reproduce the background amplitudes.  

The coincident $p(\vec{e},e^{\prime}\pi)$ cross section in
the one-photon-exchange-approximation can be written
as~\cite{obs}:
\begin{eqnarray}
\frac{d\sigma}{d\omega d\Omega_{\rm e} d\Omega^{\rm cm}_{\pi}}
    = \Gamma_{\rm v}~\sigma_{\rm h}(\theta,\phi)   \\ \nonumber
	\sigma_{\rm h}(\theta,\phi)=
	\sigma_{\rm T} + \varepsilon \sigma_{\rm L} +
	\sqrt{2\varepsilon(1+\varepsilon)} \sigma_{\rm TL} \cos\phi\\ \nonumber
 	+ \varepsilon \sigma_{\rm TT}\cos 2\phi 
 + h p_{\rm e}\sqrt{2\varepsilon(1-\varepsilon)}\sigma_{\rm
TL^{\prime}}\sin \phi\quad 
\label{eq:xsection}
\end{eqnarray}
 \noindent
where $\Gamma_{\rm v}$ is the virtual photon flux, $h = \pm 1$ is
the electron helicity, $p_{\rm e}$ is the magnitude of the
longitudinal electron polarization,
$\varepsilon$ is the virtual photon  polarization,
$\theta $ and $\phi$  are the pion CM polar and azimuthal
angles relative to the momentum transfer $\vec{q}$, and
$\sigma_{\rm L}$, $\sigma_{\rm T}$, $\sigma_{\rm TL}$, and $\sigma_{\rm
TT}$ are the longitudinal, transverse, transverse-longitudinal, and
transverse-transverse interference cross sections,
respectively. Each of these partial cross sections can be written in terms of the multipoles\cite{obs}. The E2 and M1 amplitudes can be obtained from a combination of $\sigma_{\rm T}$ and $\sigma_{\rm TT}$ as was done in photo-production (the polarized photon asymmetry =$\sigma_{\rm TT} / \sigma_{\rm T}$). In the  approximation that only s and p wave pions are produced they can be written as\cite{obs}:
\begin{eqnarray}
	\sigma_{\rm T}(\theta) &=&
		A_{T}  +B_{T} \cos\theta + C_{T} \cos^{2}\theta \\
	\sigma_{\rm TT}(\theta) &=& \sin^{2}\theta A_{TT} \nonumber  \\
	A_{T} & \approx & 5/2 |M_{1+}|^{2}+ Re[M_{1+}M_{1-}^{*}-3M_{1+}E_{1+}^{*}]
 \nonumber  \\
	B_{T} & \approx & 2M_{1+}E_{0+}^{*} \nonumber  \\
      C_{T}& \approx & -3/2|M_{1+}|^{2}+Re[9M_{1+}E_{1+}^{*}-3M_{1+}M_{1-}^{*}] 
\nonumber  \\
      A_{TT} & \approx & -1/2|M_{1+}|^{2} - Re[M_{1+}E_{1+}^{*} 
+M_{1+}M_{1-}^{*}] \nonumber 
\label{eq:T}
\end{eqnarray}
\noindent
where the pion production multipole amplitudes
are denoted by $M_{l\pm }$, $E_{l\pm }$, and $L_{l\pm }$,
indicating their character (magnetic, electric, or
longitudinal), and their total angular momentum ($J$=$l\pm
1/2$)\cite{mult}. The expressions for $A_{T},\\ B_{T}, C_{T}$ and $A_{TT}$ are in the truncated multipole approximation where it is assumed that
only terms which interfere with the dominant magnetic dipole amplitude $M_{1+}$ are kept. The exact formulas without this approximation
can be found in \cite{obs}. In this approximation the longitudinal cross section $\sigma_{L}$ = 0. In model calculations\cite{SL,MAID,KY,Az,SAID} this approximation is not made and significant deviations from 
the truncated multipole approximation occur. 

The C2 amplitude can be obtained from $\sigma_{\rm TL}$. An example of this is presented in Fig.4 which shows the Bates data\cite{Mertz} and the difference between the cross section with and without the quadrupole C2 amplitude calculated with the MAID model\cite{MAID}. The sensitivity is quite large, again indicating  a significant d state component in the $N, \Delta$ system. As can be seen from a comparison of Figs. 3 and 4, there is far more sensitivity to the C2 as compared to the E2 amplitude, despite the fact that they are both only a few \% of the M1 amplitude. The reason for this difference lies in the fact that in the longitudinal amplitude the C2 is a leading term, whereas in the transverse amplitude the E2 occurs in a linear combination with the M1 amplitude. 

The TL$^{\prime}$ and the TL (transverse-longitudinal) response functions
are the real and imaginary parts of the same combination of interference
multipole amplitudes. Again assuming that only s and p wave pions are produced they can be written as\cite{obs}:
\begin{eqnarray}
	\sigma_{\rm TL}(\theta) &=&
		- \sin\theta {\rm Re}[ A_{\rm TL} +B_{\rm TL}
\cos\theta]
	\\
	\sigma_{\rm TL'}(\theta) &=&
		\sin\theta {\rm Im}[ A_{\rm TL} +B_{\rm TL}
\cos\theta]
	\nonumber \\
	A_{\rm TL} &\approx& -L_{0+}^{*}M_{1+} \nonumber
\\
	B_{\rm TL} &\approx& -6L_{1+}^{*}M_{1+}\nonumber
\label{eq:TL}
\end{eqnarray}
where the last two approximations are for the truncated multipole approximation.
\begin{figure}
\resizebox{0.45\textwidth}{!}
%\vspace{5cm}
{\includegraphics{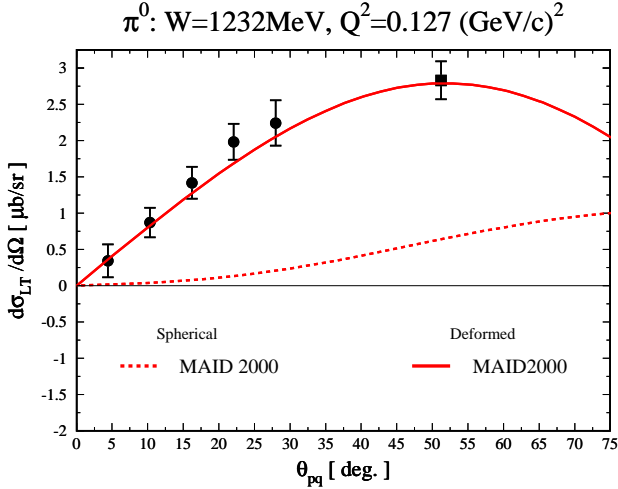}}
\caption{The differential cross section for  the $e p \rightarrow e^{'} \pi^{0} p$  reaction plotted versus $\theta_{\rm{pq}}$, the CM angle of the proton relative to the momentum transfer $\vec{q}$ (the pion CM angle $ = 180^{0} - \theta_{\rm{pq}}$)\cite{Mertz,Kunz}. 
The curves are from the MAID model\cite{MAID} with and without the C2 contribution.}
\label{sigTL}
\end{figure}

\begin{figure}
\resizebox{0.5\textwidth}{!}
%\vspace{5cm}
{\includegraphics{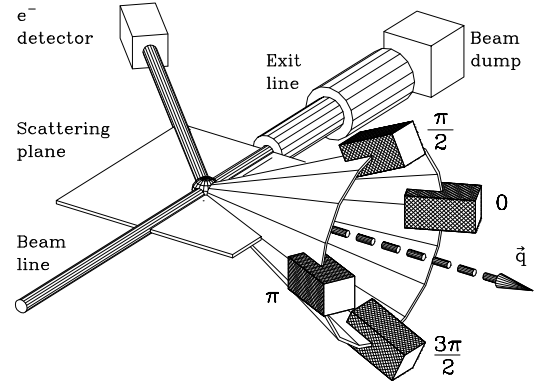}}
\caption{A schematic diagram of the out-of-plane spectrometer system (OOPS) at Bates. See text for discussion}
\label{fig:oops}
\end{figure}

The Bates out-of-plane spectrometer system (OOPS)\cite{OOPS} which was designed and built in order to exploit the $\phi$ dependence shown in Eq.~\ref{eq:xsection} is shown schematically in Fig.~\ref{fig:oops}. It consists of an electron spectrometer used in conjunction with four relatively light spectrometers which can be deployed at a fixed polar angle $\theta_{hq}$, relative to the momentum transfer $\vec{q}$ to detect the charged, emitted hadron $(p,\pi^{+})$. By deploying multiple (3 or 4) spectrometers at different azimuthal angles $\phi$, the combination of $\sigma_{0}=\sigma_{T}+ \epsilon \sigma_{L}, \sigma_{TT}$ and $\sigma_{TL}$ can be simultaneously measured in one run, which reduces the systematic errors caused by luminosity measurement errors. Furthermore, the geometry is optimized to measure relatively small relative magnitudes of $\sigma_{TL}/\sigma_{0}$ and $\sigma_{TT}/\sigma_{0}$. The combination of high luminosity and small systematic errors, allows precise measurements to be performed. Furthermore, when polarized electron beams are employed, measurements of the fifth structure function $\sigma_{TL^{'}}$, which require out-of-plane hadron detection, become possible. 

\begin{figure}
\resizebox{0.5\textwidth}{!}
%\vspace{5cm}
{\includegraphics{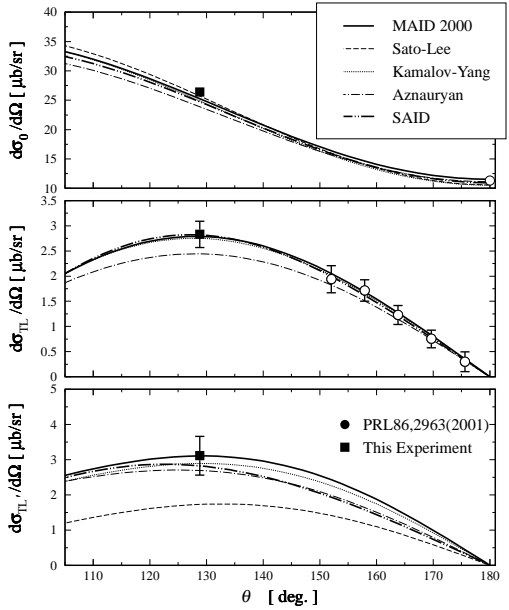}}
\caption{
Cross sections for the $p(\vec{e}, e' p)\pi^{0}$ reaction for
$W = 1232$ MeV, $Q^{2} = 0.127$ (GeV/c)$^{2}$ plotted
versus $\theta$. The top panel is for $\sigma_{0} = \sigma_{\rm T} +
\epsilon \sigma_{\rm L}$. The middle panel is for $\sigma_{\rm TL}$
and the bottom panel is for $\sigma_{\rm TL^{\prime}}$.
The curves are MAID\cite{MAID} (solid), Sato-Lee\cite{SL} (dashed),
Kamalov et al\cite{KY} (dotted), dispersion theory\cite{Az} (dot-dashed), 
and empirical multipole fit to
previous pion electroproduction data(SAID)\cite{SAID} (long-dashed).}
\label{OOPS.1232}
\end{figure}

Several rounds of experiments have been carried out at Bates with the OOPS apparatus\cite{Mertz,Kunz}. Experiments have been carried out at $Q^{2}= 0.127 GeV^{2}$ over a range of CM energies W, below, on, and above the $\Delta$ resonance energy W = 1232 MeV. For brevity only the results at the $\Delta$ peak are presented in Fig.~\ref{OOPS.1232}. The experimental results are compared to calculations\cite{SL,MAID,KY,Az,SAID}. The most ambitious
calculations are the  dynamical Sato-Lee model\cite{SL} and a dispersion
relation calculation\cite{Az}. The Sato-Lee model calculates all of the 
multipoles and $\pi-N$ scattering from dynamical equations.
Dispersion relation calculations have previously provided good agreement with photo-pion production 
data\cite{Mainz-dispersion}. Unfortunately, neither of these calculations 
agrees with the Bates data. The Sato-Lee model\cite{SL} agrees  for 
the unpolarized cross sections $\sigma_0=\sigma_{\rm T}+\epsilon \sigma_{\rm L}$ but 
is in strong disagreement with the measurements of $\sigma_{\rm TL}$ 
and $\sigma_{\rm TL^{\prime}}$. The dispersion relations calculation \cite{Az} agrees 
with some of the Bates data but disagrees with the measurement of $\sigma_{0}$ at
W = 1170 MeV(not shown here) and with $\sigma_{\rm TL}$ measurements. 
Only the two most empirical models\cite{MAID,SAID} give reasonable overall fits to all of the Bates data. The Mainz Unitary Model (MAID) is a flexible way to fit observed cross 
sections as a function of $Q^{2}$\cite{MAID}. 
It incorporates Breit-Wigner resonant terms, Born
terms,  higher $N^{*}$ resonances, and is unitarized using empirical
$\pi-$N phase shifts. The fitted parameters of the model include a range of 
data\cite{MAID}. The SAID calculation\cite{SAID} is an empirical multipole fit to
previous electropion production data.
The Kamalov-Yang model\cite{KY} includes dynamics for the resonant channels 
and uses the background amplitudes of  the MAID model. This model is in
reasonable agreement with most of the Bates data with the
exception of the unpolarized cross section $\sigma_{0}$ at $W$=1170 MeV(not shown here).

The Sato-Lee dynamical model\cite{SL} predicts that the pion 
cloud is the dominant contribution to the quadrupole amplitudes at low 
values of $Q^{2}$. Unfortunately, this model is not in agreement with our 
data but showed much better predictions of the recently reported result
from the CLAS detector at JLab for the $p(\vec{e}, e' p)\pi^{0}$  reaction
in the $\Delta$ region for $Q^{2}$ from 0.4 to 1.8 (GeV/c)$^{2}$\cite{Joo}.
This seems to indicate that the dominant meson cloud contribution, which
is predicted to be a maximum near our values of $Q^{2}$, is not
quantitatively correct. n.

Recently, a measurement of $A_{\rm TL^{\prime}}$ for the
$p(\vec{e}, e' p)\pi^{0}$  reaction in
the $\Delta$ region was performed at Mainz\cite{MainzTL'}. The kinematics
include a range of $Q^{2}$ values from 0.17 to 0.26 (GeV/c)$^{2}$ and
backward $\theta$ angles. These data were compared to several models\cite{SL,MAID,KY} which all disagreed with the data. The results of the MAID calculation had to be multiplied by 0.75 to agree with the experiment\cite{MainzTL'}. In comparison with the Bates  $\sigma_{TL^{'}}$ data\cite{Kunz}, if one multiplies the MAID results by the same factor these data are still in agreement at W=1170 MeV, but at W=1232 MeV they do not agree with a discrepancy of $1.4\sigma$. Therefore a reduction of 25\% in the MAID predictions for $\sigma_{TL^{'}}$  does not seriously effect the agreement with the Bates experiment.

It is of interest to compare the TL and TL$^{\prime}$ results presented here
with those of the recoil polarizations which are
proportional to the real and imaginary parts of interference
multipole amplitudes. For the $p(\vec{e},e' \vec{p})\pi^{0}$
channel the outgoing proton polarizations have been observed in parallel kinematics (the protons emitted along $\vec{q}$ or $\theta = 180^{\circ}$)
\cite{warren,Mpolp}. For this case the observable amplitudes are\cite{obs}:
\begin{eqnarray}
\sigma_{0}\: p_{x}&\propto&Re[A_{TL}^{x}]\\ \nonumber
\sigma_{0}\:p_{y}&\propto& Im[B_{TL}^{y}]\\ \nonumber
\sigma_{0}\:p_{z}&\propto& Re[C_{TT}^{z}]\\ \nonumber
A_{TL}^{x}\approx
B_{TL}^{y}&\approx&(4L_{1+}^{*} -L_{0+}^{*}+ L_{1-}^{*}) M_{1+} \\ \nonumber
C_{TT}^{z}&\approx& \mid M_{1+}\mid^{2}+Re[(6E_{1+}^{*}-2E_{0+}^{*})M_{1+}]
\label{eq:Pn}
\end{eqnarray}
\noindent
where $\sigma_{0}$ is the unpolarized cross section and the constants  of proportionality contain only kinematic factors. The formulas for $A_{TL}^{x}, B_{TL}^{y},C_{TT}^{z}$ assume s and p wave pions are produced and are in the truncated multipole approximation. This shows both the similarity and detailed difference
between a measurement of TL and TL$^{\prime}$ and the recoil polarizations.  In the published papers the data were compared to the MAID model which is  not in good agreement with the data \cite{warren,Mpolp}.
At the present time we do not have sufficient data to pin down the
multipoles which are responsible for this difference. 

The EMR and CMR ratios as a function of $Q^{2}$  were presented by Tiator at this workshop\cite{Tiator}. For completeness these are included here as Figs. 7 and 8. The results have provided us with a reasonably consistent overall picture of the EMR and CMR ratios. Although not presented here the pion cloud models\cite{SL,KY} are not in agreement with the data shown in Figs. 7 and 8 (see e.g. the figure presented by Burkert\cite{VB}). This agrees with the observation that this model\cite{SL} does not agree with the data near $Q^{2}= 0.125 GeV^{2}$ as discussed above. In this connection it should be noted that there is a paucity of data between the photon point and $Q^{2}= 0.4 GeV^{2}$. 
We are planning to perform an experiment at Mainz to study this region where the pion cloud effects are predicted to be large\cite{SL,KY} to test the idea of pion cloud dominance of the E2 and C2 transition amplitudes at low $Q^{2}$. 

At  large (asymptotic) values of $Q^{2}$ it is predicted that due to helicity conservation $EMR= E_{1+}/M_{1+} \rightarrow 1$ and that $CMR= S_{1+}/M_{1+} \rightarrow  constant$\cite{asym}. At the highest measured values of $Q^{2} = 4 GeV^{2}$ it is clear from Figs. 7 and 8 that the asymptotic QCD region has not been reached. 

\begin{figure*}
\begin{center}
\resizebox{0.55\textwidth}{!}
%\vspace{5cm}
{\includegraphics{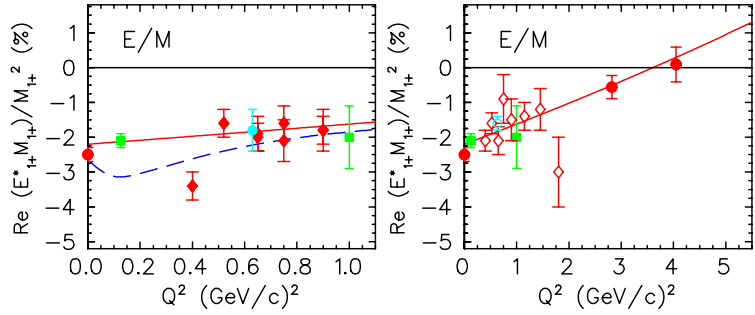}}
\caption{The $Q^{2}$ dependence of the EMR ratio\cite{Tiator}. The solid and dotted curves are for MAID\cite{MAID} and a dynamical model\cite{KY}. In the left panel the photon point is from Mainz\cite{Mainz}, the point at $Q^{2}=0.125 GeV^{2}$ is from Bates\cite{Mertz}, the circle at $Q^{2}=0.63 GeV^{2}$ is from Bonn\cite{Bonn}, the point at $Q^{2}= 1.0 GeV^{2}$ is the MAID analysis\cite{Tiator} of the Hall A JLab data\cite{HallA}, and the points from $Q^{2} = 0.4$ through $0.9 GeV^{2}$ are from the truncated multipole analysis of the Hall B JLab data\cite{Joo}. Note that for this latter set of data there are several points at the same value of $Q^{2}$ which correspond to data taken at two different beam energies. The dispersion of these points shows possible systematic errors in the data and analysis. The  right hand panel shows the MAID analysis\cite{Tiator} of the same data points as in the left panel plus the highest $Q^{2}$ points\cite{Stoler}}.
\label{fig:E2}
\end{center}
\end{figure*}
\begin{figure*}
\begin{center}
\resizebox{0.55\textwidth}{!}
%\vspace{5cm}
{\includegraphics{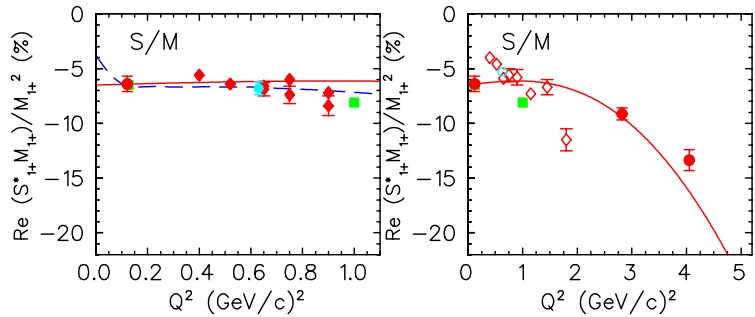}}
\caption{The $Q^{2}$ dependence of the CMR ratio\cite{Tiator}. The points at $Q^{2}=0.125 GeV^{2}$ are from the Mainz recoil polarization measurement(circle)\cite{Mpolp} and the partially hidden square from Bates\cite{Mertz}. All the other points are from the same source as in Fig.7.}
\label{fig:C2}
\end{center}
\end{figure*}

\section{What Are the Requirements for a Model Independent Data Analysis?}
\label{sec:model_independent}

It is interesting to consider how many data are required to perform a complete, model independent, multipole analysis. This can be illustrated in the approximation where it is assumed that only s and p wave pions are emitted (the discussion can be easily generalized without this assumption). In this case there are 7 multipoles for electroproduction ($E_{0+}, E_{1+}, M_{1\pm}, L_{0+}, L_{1\pm}$). Since these are complex they represent 14 numbers. However, an overall phase is irrelevant, so this leaves 13 numbers to be determined at each value of $Q^{2}$ and W. By counting the observables in Eqs. 2 and 3  it can be seen that for for polarized electrons and unpolarized  targets there are 8 observables, namely A,B,C, $A_{TT},Re[A_{TL}], Im[A_{TL}], Re[B_{TL}], Im[B_{TL}]$ in Eqs 2 and 3. This is the number we have obtained at Bates (including data presently being analyzed)at $Q^{2} = 0.127 GeV^{2}$. By adding recoil polarization observables in parallel kinematics, 3 more numbers are measured(Eq. 4). This provides a stringent test of the reaction models but not enough to make a model independent analysis. By measuring recoil polarization away from the forward direction the remainder can be measured. It is of interest to compare this to the photon experiments which have been carried out with polarized photons and unpolarized targets. There are 4 transverse multipoles which make 8-1 = 7 numbers to determine. The actual experiments\cite{Mainz,LEGS} determined 4 of these (A,B,C and $A_{TT}$ of Eq. 2). So even these data are not yet sufficient for a model independent analysis, and polarized target data is required to complete this task. Such experiments are underway at LEGS in Brookhaven and are being planned at Mainz.

\section{Conclusions}
\label{sec:conclusions}
In conclusion, our data are consistent with a  deviation of the nucleon and $\Delta$ from spherical symmetry. We are making rapid progress towards making a more quantitative measurement of this effect. The errors are primarily in the model extraction of the deformation. We also are making rapid progress towards stringently testing the reaction models, which means pinning down the resonant and background amplitudes. The latter effort consists of measurements of the fifth structure function $\sigma_{TL^{'}}$ and of the recoil polarizations. In addition we are also making progress towards a sufficient data-base to approach making model independent analyses. The interesting question of whether the main mechanism for the deviation from spherical symmetry has its origin in the spontaneous breaking of chiral symmetry in QCD with the resulting long range contribution from the pion cloud, needs further experimental and theoretical work. 

\section{Acknowledgements}
  I want to thank the organizers, O. Benhar, A. Fabrocini, and R. Schiavilla for their invitation to a stimulating workshop. I would like to thank my OOPS colleagues for their collaboration, and for allowing me to use data prior to publication. I also thank L. Tiator for discussions during the workshop and for permission to use  Figs.7 and 8.  For their helpful discussions of the manuscript I would like to thank V.R.Brown,B. Holstein, T.-S.~H.~Lee, U.G.Mei{\ss}ner,  S.Stave, I. Nakagawa, C.Vellidis.


\begin{thebibliography}{99}
\bibitem{mult} The relation between the transition electric and magnetic multipoles and the pion multipoles is $M1= M_{1+}, E2= E_{1+}, C2=S_{1+}$ where the scalar multipoles $S_{l\pm}$ are simply related to the longitudinal multipoles (used in the text) as $\omega S_{l\pm}= q L_{l\pm}$ where $\omega$ and q are the center of mass energy and three momentum transfers respectively.
\end{thebibliography}
\end{document}